\documentclass[%
 reprint,
 amsmath,amssymb,
 aps,
 nofootinbib,
]{revtex4-2}
\usepackage[english]{babel}
\usepackage{enumitem}
\usepackage{hyperref}
\usepackage{xcolor}
\usepackage{braket}
\usepackage{subcaption}
\usepackage{pdfcomment}
\usepackage{todonotes}
\usepackage{nicematrix}
\usepackage{siunitx}[=v2]
\usepackage{graphicx} 
\usepackage[nolist,nohyperlinks]{acronym}
\usepackage{amsmath, amsthm, amssymb, amsbsy, mathtools, mathalpha}

\usepackage{graphicx}
\usepackage{dcolumn}
\usepackage{bm}
\usepackage{hyperref}
\usepackage{color, units}
\usepackage[normalem]{ulem}

\usepackage{orcidlink}
\usepackage{xspace}
\usepackage{comment}
\usepackage{xcolor}

\graphicspath{{figures/}}

\newcommand{\avi}[1]{}
\newcommand{\RM}[1]{}
\newcommand{\kl}[1]{}
\newcommand{\KJ}[1]{}
\newcommand{\NC}[1]{}
\newcommand{\pmr}[1]{}
\newcommand{\jliu}[1]{}
\newcommand{\YL}[1]{}

\renewcommand{\d}{{\bf d}}

\newcommand{\bepsilon}{\mbox{\boldmath $\epsilon$}}

\newcommand{\blambda}{\mbox{\boldmath $\lambda$}}
\newcommand{\bSigma}{\mbox{\boldmath $\Sigma$}}
\renewcommand{\S}{{\bf S}}
\newcommand{\A}{{\bf A}}
\newcommand{\C}{{\bf C}}
\newcommand{\G}{{\bf G}}

\newcommand{\U}{{\bf U}}
\newcommand{\Y}{{\bf Y}}
\newcommand{\Z}{{\bf Z}}

\newcommand{\eqq}{\ensuremath{\!\mathrel{\scalebox{0.7}{=}}\!}}
\newcommand{\tmes}{\ensuremath{{\mkern-2mu\times\mkern-2mu}}}
\newcommand{\ex}[1]{\ensuremath{\tmes10^{\text{-}#1}}}

\newcommand{\tr}{\mbox{tr}}
\newcommand{\I}{\mbox{{\bf I}}}
\newcommand{\e}{{\rm e}}
\newcommand{\M}{{\bf M}}

\DeclareSIUnit{\kilo}{k}

\begin{document}

\preprint{APS/123-QED}

\title{Bayesian nonparametric estimation of correlated gravitational wave detector network noise using matrix-gamma process priors
}
\author{Yixuan Liu$^{1}$\orcidlink{0009-0000-1528-4016}}
\author{Renate Meyer$^{2}$\orcidlink{0000-0003-0268-8569}}
\author{Nelson Christensen$^{3}$\orcidlink{0000-0002-6870-4202}}
\author{Jeung Eun Lee$^{2}$\orcidlink{0000-0003-4993-4683}}
\author{Jianan Liu$^{2}$\orcidlink{0009-0007-4718-1215}}
\author{Patricio Maturana-Russel$^{2,4}$\orcidlink{0000-0002-5211-9818}}
\author{Avi Vajpeyi$^{2}$\orcidlink{0000-0002-4146-1132}}

\affiliation{$^{1}$MRC Biostatistics Unit, University of Cambridge, Cambridge, UK}
\affiliation{$^{2}$Department of Statistics, University of  Auckland, 38 Princes St, Auckland, New Zealand}%
\affiliation{$^{3}$Universit\'e C$\hat{o}$te d’Azur, Observatoire C$\hat{o}$te d’Azur, CNRS, Artemis, 06304 Nice, France}
\affiliation{$^{4}$Department of Mathematical Sciences, Auckland University of Technology, Auckland, New Zealand}



\begin{abstract}
This paper addresses the important problem of estimating the noise spectral density of next-generation gravitational-wave detectors, such as LISA and the Einstein Telescope (ET), where cross-channel correlations must be accounted for to avoid biased  parameter estimation of gravitational-wave signals. Unlike approaches that estimate test-mass and optical-metrology-system noise separately at the single-link level and then map them to the Time-Delay Interferometry (TDI) channels through known transfer functions, we develop a Bayesian nonparametric method that directly estimates the spectral density matrix of the XYZ channels, thereby accommodating additional sources of uncertainty. Our approach combines a flexible matrix-gamma process prior on the matrix-valued coefficients of a Bernstein polynomial basis expansion with a blocked multivariate Whittle likelihood. The prior guarantees Hermitian positive definiteness of the spectral estimate at every frequency.
To avoid reversible-jump methods, we use an adaptive Markov chain Monte Carlo (MCMC) algorithm for posterior sampling. The proposed framework can also be used to correct misspecified parametric noise models. Results from a simulation study and simulated correlated-noise data for both LISA and ET demonstrate the effectiveness of the proposed method. 
\end{abstract}

\maketitle

\begin{acronym}
    \acro{GW}[GW]{gravitational-wave}
    \acro{GWs}[GWs]{gravitational-waves}
    \acro{ET}[ET]{Einstein Telescope}
    \acro{LISA}[LISA]{Laser Interferometer Space Antenna}
    \acro{CE}[CE]{Cosmic Explorer}
    \acro{PE}[PE]{parameter estimation}
    \acro{PSD}[PSD]{Power Spectral Density}
    \acro{DFT}[DFT]{Discrete Fourier Transform}
    \acro{SNR}[SNR]{Signal-to-Noise Ratio}
    \acro{BBH}[BBH]{binary black hole}
    \acro{VI}[VI]{variational inference}
    \acro{KL}[KL]{Kullback-Leibler}
    \acro{MCMC}[MCMC]{Markov chain Monte Carlo}
    \acro{LVK}[LVK]{LIGO--VIRGO--KAGRA}
    \acro{SGVB}[SGVB]{stochastic gradient variational Bayes}
\end{acronym}

\section{Introduction}
Accurate noise characterization is a central challenge for next-generation
gravitational-wave detectors. A misspecified power spectral density (PSD) can
bias signal parameters, affect model selection, and lead to miscalibrated
uncertainties~\cite{LittenbergTysonB.2015Bifs,ChatziioannouKaterina2019Nsem}.
Both the Einstein Telescope (ET) and the Laser
Interferometer Space Antenna (LISA) will produce multiple, coupled data
streams, and the scientific return from each depends on how well the spectral
density of the instrumental noise can be estimated across channels. For
multichannel data, the noise is characterized by the spectral density matrix,
whose diagonal entries are the channel PSDs and whose off-diagonal entries
describe the frequency-dependent correlations between channels. A suitable
model must capture this frequency dependence while ensuring that the inferred
spectral density matrix remains Hermitian positive definite at every frequency.
Unlike the
current LVK network, where spatial separation makes channel independence a
reasonable assumption, ET and LISA operate in regimes where cross-channel
correlations are physically expected and cannot be safely ignored in
gravitational-wave parameter inference.

In ET, the triangular configuration and the proximity of the test masses lead 
to correlated magnetic, seismic, and Newtonian noise contributions~\cite{Thrane:2013npa,Meyers:2020qrb,Janssens:2022tdj,Janssens:2022xmo,Janssens:2024jln,schillings2026numericalframeworknewtoniannoiseestimation}. 
So far, most spectral density estimation techniques developed for ET  have either ignored the presence of correlation or tried to mitigate their impact by subtracting the correlations \cite{RevModPhys.94.025001,Badaracco}.
 However, several papers have pointed out the necessity of taking the instrumental noise correlations into account to avoid biasing signal parameter estimates or falsely claiming overly confident uncertainties \cite{Janssensmagneticnoise,Janssensnewtonianseismic,JanssensKamiel2023Ffps,PhysRevD.110.104060}. 
In LISA, the measured channels are related through 
link-level test-mass and optical-metrology noise processes and their transfer 
functions \cite{PhysRevD.108.082004}. 
The main approach for estimating the instrumental noise of the correlated time delay interferometry (TDI) XYZ channels is to estimate the dominant secondary test mass and optical metrology system noise at the single link level, assuming independent noises in the six movable optical sub-assemblies (MOSAs) and known transfer functions that map from the six single links to the three XYZ channels \cite{BaghiQuentin2023Ugbf,SantiniAlessandro2025AfGa}. However, some uncertainty remains \cite{PhysRevD.108.082004,PhysRevD.109.042001}, e.g., due to TDI ranging, i.e., uncertainty in measurements of the distances between spacecraft, and tilt-to-length (TTL) coupling \cite{PhysRevD.111.043048}. Therefore, it is preferable to model the correlated noise in TDI channels directly to capture all noise contributions and correlations.
Correlated noise is also directly relevant for
stochastic-background searches, where disentangling
instrumental correlations from gravitational-wave power is part of the signal
model itself: \citet{PhysRevD.109.042001} show that allowing for realistic
uncertainty in the LISA auto- and cross-spectral densities degrades the
achievable constraints on a stochastic background by one to two orders of
magnitude, and that avoiding this degradation would require the noise to be
known at the sub-percent level.
Estimating the noise spectral density matrix for ET or LISA in a way that is
useful for downstream gravitational-wave inference requires four things at once:
flexibility across the full $d\times d$ matrix, so that both localized spectral
features and cross-channel structure can be recovered without prescribing a
functional form; a guarantee that the estimate is Hermitian positive definite at
every frequency, so that it can be inverted inside a likelihood; honest
posterior uncertainty, so that the noise uncertainty can be propagated into the
signal parameters rather than fixed; and enough computational tractability to
handle the very long records that both instruments will produce. Existing
approaches meet some of these but not all.
Welch-type estimates \cite{WelchP.1967Tuof} are flexible and cheap but return a
single fixed spectral density, so they supply no uncertainty to propagate.
Restrictive parametric models such as \cite{PhysRevD.103.103529,Tang2025} are
computationally efficient and quantify uncertainty within their own
parameterization, but may miss localized spectral features or cross-spectral
structure that the parameterization does not anticipate. Spline-based
perturbations of a design noise curve \cite{PhysRevD.109.042001} relax this by
allowing smooth fractional deviations in both the auto- and cross-spectra, but
they are tied to a reference model and, as those authors note, do not by
construction keep the resulting spectral density matrix positive definite.
Variational methods that approximate the likelihood and avoid Markov chain
Monte Carlo (MCMC) methods can scale to long time series \cite{Blei03042017}; in
particular, the stochastic gradient variational Bayes approximation
developed by \cite{liu2024} is flexible and fast and
often gives reasonably accurate point estimates, but tends to underestimate
posterior uncertainty -- which is precisely the quantity one wants to propagate.
Closest to this work, \cite{avi:2026} model the inverse spectral density matrix
through a frequency-dependent Cholesky decomposition, using Bayesian P-splines
and No-U-Turn sampling to estimate LISA TDI spectra; the Cholesky construction
guarantees positive definiteness, and here we instead place a stochastic-process
prior directly on the matrix-valued coefficients of a spectral basis expansion.
What is missing is a method that is nonparametric over the whole spectral
density matrix, positive definite by construction, and sampled from the exact
posterior rather than an approximation to it. 
Such a method is necessarily more computationally demanding than the alternatives described above, but it provides a valuable benchmark against which the accuracy of the faster approximation methods can be evaluated.


To model the multichannel noise efficiently while preserving its spectral dependencies, we  segment the time series and average the resulting periodogram matrices to obtain a blocked
Whittle likelihood in complex-Wishart form. 
This reduces the effective data 
volume while retaining the full cross-spectral structure of the multichannel 
noise without requiring a low-dimensional parametric form. We combine this 
likelihood with a proper nonparametric matrix-gamma process prior on the space of positive definite power spectral density (PSD) matrices with a potentially infinite number of basis functions
\cite{MeierAlexander2020Bnao}. This matrix-gamma process prior is a generalization of the Bernstein-Dirichlet 
process prior for univariate spectral densities to the multivariate
setting. The prior avoids reversible-jump MCMC by representing the matrix-valued basis  coefficients as values of a cumulative distribution function and placing a nonparametric
matrix-gamma process prior on that distribution. A key theoretical property is 
that the posterior is guaranteed to concentrate around the true spectral 
density and forgets the nonparametric prior as the data record grows \cite{DiaconisPersi1986OtCo}. Thus, the 
inferred spectrum is eventually driven by the data rather than the prior. The model will be referred to as VNP (vector nonparametric) in the following. 

Nonparametric methods are very flexible but usually less efficient than parametric methods when the parametric model is well specified. To answer the question of whether it is possible to improve the performance of nonparametric PSD estimation using a well-specified parametric ``working'' model, we introduce in Section~\ref{sec:methodology} the VNP-P (vector nonparametrically corrected parametric) model. This can be regarded as correcting a parametric PSD by the introduction of a nonparametric component. 
Here a fitted parametric 
spectral-density matrix captures the broad spectral shape, and the 
nonparametric component corrects residual structure such as
 mis-modeled peaks or cross-channel correlations. We assess, rather than assume,
this benefit: the simulation study includes a case in which the working model
captures useful structure and a case in which it is misspecified. We validate the nonparametric methods
 on controlled simulations and on LISA-like and  ET-like
examples. The goals of this work are to show accuracy 
and to establish this MCMC sampler as a potential reference against which faster 
methods that use likelihood approximations, such as SGVB \cite{liu2024} and coarsened likelihoods
\cite{FlaugerRaphael2021Iroa}, or deep-learning point estimators 
\cite{PhysRevLett.130.171403} can be assessed.

The paper is organized as follows. Section~\ref{sec:methodology} introduces the 
notation, the blocked Whittle likelihood in complex-Wishart form, and the 
matrix-gamma process Bernstein prior. Section~\ref{sec:simulation} presents 
simulation studies of estimation accuracy. 
 Section~\ref{subsec:LISA} applies the VNP method to 
the LISA XYZ channels.
Section~\ref{subsec:ET}  compares the 
nonparametric VNP and parametric variant VNP-P on simulated ET-like data in terms of 
accuracy and computation time over a restricted frequency range.
Section~\ref{sec:discussion} summarizes 
the conclusions and discusses how this nonparametric MCMC benchmark can be used to validate 
faster approximate methods.

\section{Methodology}\label{sec:methodology}
\subsection{Likelihood}\label{sec:likelihood}


Assume we are given $n$ gravitational-wave observations $\Z_t=(Z_1,\ldots,Z_d)^\intercal$ from $d$ channels, for $t=1,\ldots,n$. Let
$\Z=(\Z_1,\ldots,\Z_n)^\intercal \in \mathbb{R}^{n\times d}$
denote the corresponding data matrix. We assume that $\{\Z_t\}_{t=1}^n$ is a $d$-dimensional, stationary, mean-zero time series sampled at intervals $\Delta_t=1/(2f_{\mathrm{Ny}})$, where $f_{\mathrm{Ny}}$ denotes the Nyquist frequency. The resulting observation period is $T=n\Delta_t$. Hence, the frequency resolution is
\begin{align}
\Delta_f=\frac{1}{n\Delta_t}=\frac{1}{T}.
\end{align}
The \ac{DFT} of $\Z$ at frequency $f_k=k\Delta_f=k/T$ is defined by
\begin{align}
\d(f_k) = \sqrt{\frac{\Delta_t}{n}}\sum_{t=1}^{n} \Z_t\exp \left(-2\pi i \frac{k}{n} t \right),
\end{align}
for $k=1,\ldots,N$, where $N=n/2$ if $n$ is even and $N=(n-1)/2$ if $n$ is odd.

The discrete Fourier coefficients $\d(f_k)$ are asymptotically independent and have a complex Gaussian distribution with covariance matrix given by the spectral density matrix $\S(f_k)$,  the Fourier transform of the autocovariance function. 
The well-known multivariate Whittle likelihood approximation uses this asymptotically independent complex Gaussian density to define a pseudo-likelihood in the frequency domain:
\begin{align}\label{eq:Whittle likelihood}
 \mathcal{L}(\d|\S) &\propto  \prod_{k=1}^{N} \left|\S(f_k)\right|^{-1} \e^{-\d(f_k)^* \S(f_k)^{-1} \d(f_k)},              
\end{align}
where $\d(f_k)^*$ denotes the conjugate transpose of the $\d(f_k)$, $\left|\A\right|$ denotes the determinant of a matrix $\A$,  and $\S(f_k)$ is a $d \times d$ Hermitian positive definite spectral density matrix at each $f_k$. 
By using the  trace rotation rule for the product of  square matrices (where the trace of a matrix is the sum of its diagonal elements, i.e., $\textrm{tr}(A)=\sum_{i=1}^n a_{ii}$), namely 
$\tr(A B C)=\tr(BC A)=\tr(C A B)$, 
the Whittle likelihood can be expressed as a function of the periodogram matrix
$\I(f_k)=\d(f_k)\d(f_k)^*$ as follows
\begin{align}\label{eq:Whittle-likelihood-periodogram}
 \mathcal{L}(\d|\S) &\propto   \prod_{k=1}^{N} \left| \S(f_k)\right|^{-1} \exp\left(-\tr\left[ \S(f_k)^{-1} \I(f_k)\right]\right).
\end{align}

As is well known, the periodogram is not a consistent estimate of the spectral density matrix as its variance does not decrease with increasing sample size.
To obtain a consistent estimate, previous studies such as \cite{PhysRevD.110.104060} have employed the classical Welch estimate \cite{WelchP.1967Tuof}, which is based on averaging periodogram matrices from overlapping data segments. This PSD estimate is usually assumed fixed for the purpose of estimating the parameters of a GW signal. However, as has been shown in the context of LIGO \cite{ChatziioannouKaterina2019Nsem,LittenbergTysonB.2015Bifs}, on-source estimation is important, i.e.,  to simultaneously estimate the spectral density with the GW parameters in order to properly quantify the associated uncertainties. This is imperative for LISA, as there will not be any stretches of data that contain pure noise and that can be used to estimate the noise spectral density.
To this end, a flexible Bayesian approach is needed that can be combined, e.g., within a Gibbs sampling scheme, to simultaneously estimate GW signal parameters. One such approach based on a blocked Whittle likelihood and a variational inference approximation has been suggested in \cite{liu2024}. 
The blocked Whittle likelihood was used to make spectral density estimation possible for very long time series and the variational approach replaced MCMC sampling by running an optimization algorithm which reduced computation time.
Here, we also  segment the data into non-overlapping blocks and employ the blocked Whittle likelihood.
However, we can considerably simplify the form of the log-likelihood. It can be seen as equivalent to using as input the averaged periodogram matrix and constructing a likelihood approximation based on its asymptotic complex Wishart distribution. This will be detailed in the following, whereas in Section~\ref{sec:prior} we will describe a flexible, nonparametric prior over the space of Hermitian positive definite spectral density matrices.


In analogy to \cite{liu2024}, we divide the time series into $N_{b}$ equal-sized blocks $\Z=(\Z^{(1)},\ldots,\Z^{(N_{b})})^\intercal $ with each block a $d$-dimensional time series of each of length $n/N_{b}$. The discrete Fourier transform of each block is denoted by 
$\d^{(i)}$, $i=1,\ldots,N_{b}$. 
Assuming stationarity of the time series and independence among different blocks, the overall blocked likelihood is the product of the individual Whittle likelihoods of each block,
\begin{equation}\label{eq:block_lnl}
    \mathcal{L}_b(\d|\S) = \prod^{N_b}_{i=1} \mathcal{L}(\d^{(i)}|\S) \ .
\end{equation}
Let $N_{l}=n/(2N_{b})$ if $n/N_{b}$ is even, and $N_{l}=(n/N_{b}-1)/2$ if $n/N_{b}$ is odd. Then, the blocked Whittle likelihood can be simplified by using the  trace rotation rule to\\
$\mathcal{L}_b(\d^{(i)}|\S)$ 
\begin{eqnarray}
&\propto & \prod_{k=1}^{N_l}   \left|\S(f_k)\right|^{-N_b}  
  \exp\left(- \sum_{i=1}^{N_b}   \tr\left[\d^{(i)}(f_k)^* \S(f_k)^{-1} \d^{(i)}(f_k)\right]\right) \notag \\ 
  &\propto &  \prod_{k=1}^{N_l}    \left|\S(f_k)\right|^{-N_b}   
  \exp\left(-   \tr\left[ \S(f_k)^{-1} \sum_{i=1}^{N_b} \underbrace{\d^{(i)}(f_k)\d^{(i)}(f_k)^*}_{\I^{(i)}(f_k)} \right]\right) \notag \\
   &\propto & \prod_{k=1}^{N_l}     \left|\S(f_k)\right|^{-N_b}   
  \exp\left(-   \tr\left[ \S(f_k)^{-1} N_b\bar{\I}(f_k)\right]\right) \notag \\ 
   &\propto &  \prod_{k=1}^{N_l}    \left|\S(f_k)\right|^{-N_b}   
  \exp\left(-   \tr\left[ \S(f_k)^{-1} \Y(f_k)\right]\right)
\end{eqnarray}
where 
\begin{equation}
\displaystyle \bar{\I}(f)=\frac{1}{N_b} \sum_{i=1}^{N_b} \I^{(i)}(f)
\end{equation}
is the block-averaged periodogram (or Welch estimate of the spectral density with non-overlapping blocks) of length $N_l=n/(2N_b)$, $\I^{(i)}(f)=\d^{(i)}(f)\d^{(i)}(f)^*$ the periodogram matrix of the $i$th block at frequency $f$, and 
$\Y(f)=N_b \bar{\I}(f)$.
Therefore, $\Y(f)$ has a $d$-dimensional complex Wishart distribution\footnote{ We use the notation $\A \sim {\cal CW}_d(\M,m)$  if  the $d\times d$ Hermitian positive definite random matrix $\A$ has a complex Wishart distribution with mean $m\M$ and $m$ degrees of freedom. Its density is proportional to $\left|\M\right|^{-m}\exp(-\tr[\M^{-1}\A])$.} with $N_b$ degrees of freedom,  i.e., $\Y(f_k)\sim {\cal CW}_d(\S(f_k), N_b)$, . 
This is consistent with the independent $d$-dimensional complex Wishart distribution with one degree of freedom for each individual periodogram, i.e.,
\begin{equation}
\I^{(i)}(f_k)\sim {\cal CW}_d(\S(f_k),1). 
\end{equation}
The expression for the log-likelihood function is thus
\begin{equation}\label{eq:log-like}
    \sum_{k=1}^{N_l} -N_b\left(\log \left|\S(f_k)\right| + \tr\left[ \S(f_k)^{-1} \bar{\I}(f_k)\right]\right).
\end{equation}

The choice of the number of blocks is usually based on ensuring a reasonable frequency resolution while still enabling MCMC sampling within a reasonable time frame.

We make one further modification to the blocked Whittle log-likelihood based on the idea of \cite{Liu2023} that a nonparametric model can be made more efficient by exploiting a reasonably well-fitted parametric model, while retaining robustness through a nonparametric correction when the parametric model is misspecified.  As we will see, this improves the estimation of sharp peaks and abrupt changes in the PSD. The approach in \cite{Liu2023} makes use of a non-parametrically corrected likelihood that is based on an autoregressive model in the time domain but non-parametrically corrected in the frequency domain. This requires a toggle  between the frequency and time domain to fit parametric and nonparametric PSD simultaneously in each iteration of a blocked Gibbs sampler. 

To avoid this time-consuming simultaneous estimation, we instead proceed in two steps: estimating the parametric spectral density first by fitting a vector autoregressive (VAR) model of fixed order $p$, i.e., a VAR($p$) model, by maximizing the blocked Whittle likelihood. 
Although we use a vector autoregressive model here, any other parametric model could be used if there are reasons to believe it provides an adequate fit. The purpose of this parametric model is not to assume that the VAR model is correct. Rather, it provides a rough spectral template that can capture sharp peaks or broad spectral structure, while the nonparametric component models the remaining discrepancy. 

In a second step,  we fix the resulting parametric spectral density matrix $\S_\text{p}$ and model the spectral density matrix $\S(f_k)$ in (\ref{eq:log-like}) as the matrix geometric mean of the parametric PSD $\S_\text{p}(f_k)$ and a nonparametric PSD $\S_\text{np}(f_k)$:
\begin{equation} \label{eq:Ando}
\S= \S_\text{p}^{1/2}\left( \S_\text{p}^{-1/2} \S_\text{np} \S_\text{p}^{-1/2}\right)^{1/2} \S_\text{p}^{1/2}
\end{equation}
using the definition of the geometric mean of two positive definite matrices in \cite{AndoT.2004Gm}. 
In (\ref{eq:Ando}) we dropped the frequency argument in each matrix for ease of notation but the relationship holds pointwise in frequency.
This is the only definition in the literature that possesses the properties that one would expect from a geometric mean and that generalizes many inequalities satisfied by the geometric mean of two positive semidefinite matrices. Thus, the factor 
\begin{align}\label{eq:correction_factor}
\C=\left( \S_\text{p}^{-1/2} \S_\text{np} \S_\text{p}^{-1/2}\right)^{1/2}    
\end{align}
can be regarded as a matrix-valued correction factor. Now, we place the nonparametric prior on this correction matrix $\C$ instead of on $\S_\text{np}$ itself. The rationale is that the parametric PSD is expected to capture the main peaks of the spectral density matrix, so that the entries of the correction matrix $\C$ are smoother functions and can be more easily approximated in a nonparametric way. 

\subsection{Prior}\label{sec:prior}
The construction of a nonparametric prior on the infinite space of matrix-valued spectral density functions that are Hermitian positive definite at each frequency is not trivial. While other approaches such as \cite{KRAFTYROBERTT.2013PmWl, liu2024} smoothed the components of the Cholesky decomposition to ensure a positive definite estimator, here we suggest to use a construction that can be regarded as a generalization of the Bernstein-Dirichlet process prior \cite{Choudhuri04} for univariate spectral densities to the multivariate case. When updated with the multivariate Whittle likelihood, \cite{MeierAlexander2020Bnao} showed good asymptotic properties such as posterior consistency which guarantees that the posterior distribution will contract around the true spectral density with increasing sample size.  
We will first describe the nonparametric prior for the univariate case and then its generalization to multivariate spectral densities.

\begin{figure}
  \includegraphics[width=1\columnwidth]{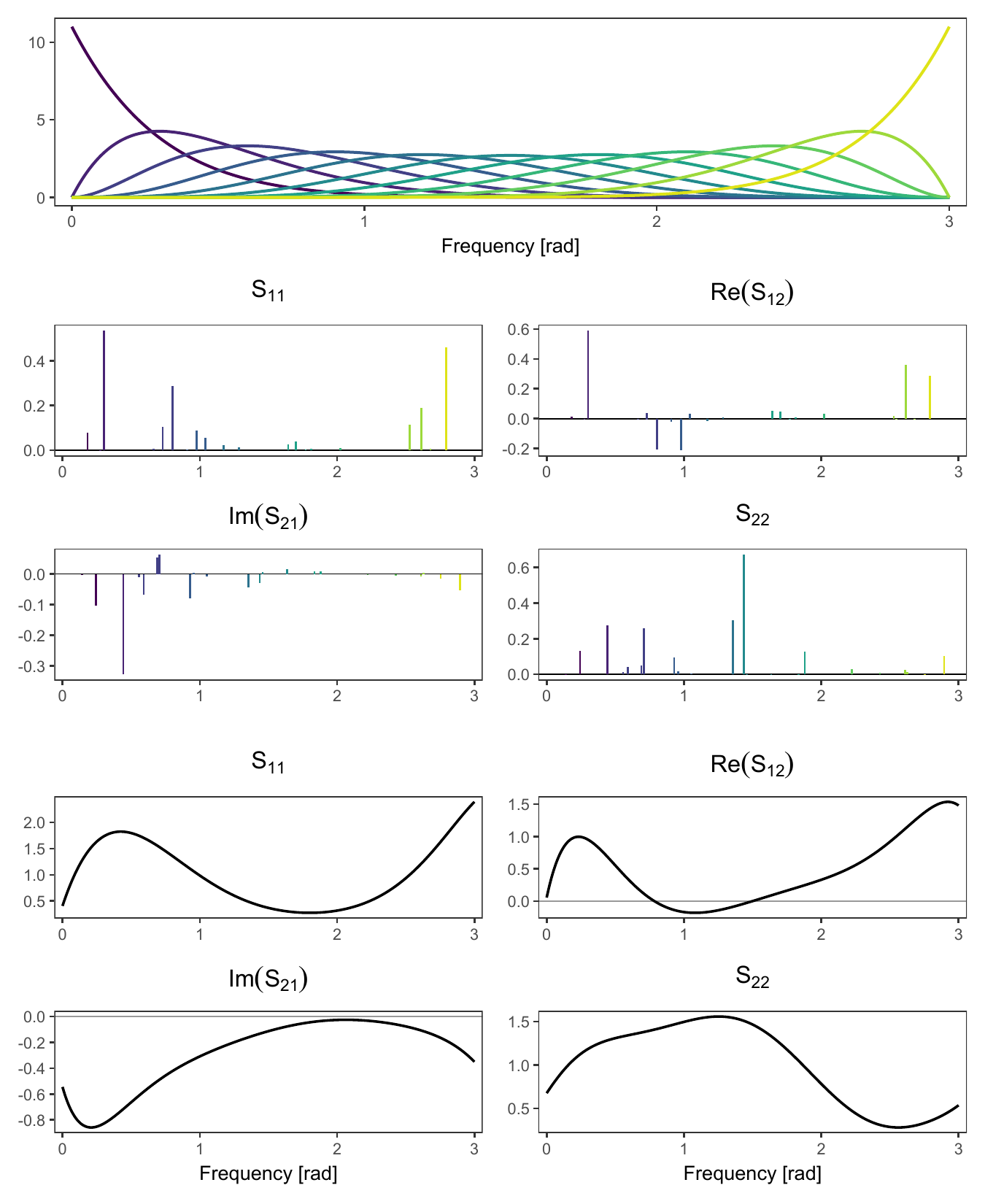}
  \caption{The  top panel displays the Bernstein polynomial basis of order $m=10$, the middle panel a draw from a matrix-gamma process prior, and the bottom panel the corresponding mixture representation of the bivariate PSD matrix with auto-spectra on the diagonal and real and imaginary parts of the cross-spectrum on the off-diagonals. 
  } 
  \label{fig:Gammaprocess}
\end{figure}

A univariate spectral density (after rescaling to [0,1]) can be approximated by a positive combination of normalized Bernstein polynomials, i.e., Beta densities on the interval $[0,1]$:
\begin{align}
S\left(\frac{1}{2\Delta_t} \omega\right)= \sum_{j=1}^m \lambda_{j,m} \beta(\omega|j,m-j+1),
\end{align}
where 
\begin{align}
\beta(\omega|j,l)= \frac{\Gamma(j+l)}{\Gamma(j)\Gamma(l)} \omega^{j-1}(1-\omega)^{l-1}
\end{align}
is the density of the Beta distributions with parameters $j$ and $l$, $\omega\in [0,1]$. The order $m$ of the polynomial approximation is a smoothing parameter, the smaller $m$, the smoother the resulting approximation. By putting a prior on $m$, we obtain a data-driven estimate of the degree of smoothness. The positive ``weights" $\lambda_{j,m}=G(j/m)-G((j-1)/m)$ are defined by the values that a probability distribution function $G$ gives to the subinterval $\bigl((j-1)/m,\; j/m\bigr]$. 
Then, a nonparametric prior is specified over the space of all distributions on $[0,1]$. Such a prior is the gamma process for $G$ with parameters $\alpha,\beta>0$ which effectively means that the increments $G(d\omega)$ have an independent Gamma$(\alpha,\beta)$ distribution.
%
%
Similarly, in the $d$-dimensional case, the
multivariate spectral density can be approximated by a mixture of Beta densities
\begin{equation}
    \S\left(\frac{1}{2\Delta_t} \omega\right)= \sum_{j=1}^m \blambda_{j,m} \beta(\omega|j,m-j+1),
\end{equation}
but the mixture weights $\blambda_{j,m}=\G(j/m)-\G((j-1)/m)$ are now Hermitian positive definite 
matrices and a Hermitian matrix-gamma process prior is constructed as in \cite{MeierAlexander2020Bnao} so that the increments $\G(dx)$ have an independent Hermitian positive definite matrix-gamma distribution \cite{MeierAlexander2020Bnao}. This is a special multivariate generalization of the univariate gamma distribution. This construction is illustrated in Figure \ref{fig:Gammaprocess} in the bivariate case for a fixed polynomial order $m=10$. Using different colours, at the top are the ten normalized Bernstein polynomials, the middle panel shows one draw from a $2\times 2$ matrix-gamma process prior whose realizations are discrete random matrices. If for each of the four panels one sums up the point masses within the subinterval $\bigl((j-1)/m,\; j/m\bigr]$, this will define the weight attached to the corresponding  Bernstein basis polynomial of the same colour.  Finally, the bottom panel displays each of the weighted averages of the Bernstein polynomials for the spectral densities on the diagonal and the real and imaginary part of the cross-spectrum on the off-diagonals.
The Hermitian positive definite matrix-gamma prior depends on three parameters $(\alpha_0,\beta_0,\bSigma_0)$ for $\alpha_0>d-1$, $\beta_0>0$ and a positive definite matrix $\bSigma_0$. Its mean is given by $(\beta_0/d)\bSigma_0$. For a noninformative choice of the hyperparameters, we set $\alpha_0=\beta_0=d$ and $\bSigma_0=10^4\mathbf{Id}_d$ proportional to the identity matrix. 
A common choice for the prior of $m$ is $p(m)\propto\exp(-c\; m\log m)$ for some positive constant $c$. In practice, we choose $c=0.01$ that is used in \cite{MeierAlexander2020Bnao, Liu2023}.

\subsection{Posterior Computation}\label{sec:posterior}
To sample from the posterior distribution using an adaptive MCMC algorithm, we make use of an infinite series representation of the Hermitian positive definite
matrix-gamma process $\G$ that depends only on iid random variables, similar to the stick-breaking representation of the Dirichlet process~\cite{Sethuraman94}
\begin{align}
\G= \sum_{j=1}^L \delta_{x_j} r_j\U_j 
\end{align}
and truncated at some large value $L$. A conservative truncation choice is $L=\max\lbrace20,(n/N_{b})^{1/3}\rbrace$, see \cite{MeierAlexander2020Bnao, Liu2023}. For details of the exact definitions of $x_j, r_j, \U_j$ see \cite{MeierAlexander2020Bnao}.

\section{Simulation Study}\label{sec:simulation}
For unsegmented multivariate time series ($N_b=1$), \cite{MeierAlexander2020Bnao}
established posterior consistency of VNP by proving that, as the sample size increases, the posterior distribution under the multivariate Whittle likelihood and matrix-Gamma process prior concentrates around the true spectral density matrix. The theoretical results were complemented by simulation studies that empirically confirmed this asymptotic behaviour for increasing time series lengths. Here, we 
illustrate the performance of VNP and VNP-P for large time series with {\em increasing block lengths}. We simulated 500 instances of two-dimensional time series of length \num{819200} from a vector autoregressive model of order 2, VAR(2), and a vector moving average process of order 1, VMA(1), for which the true spectral densities are known and can be used to compare the accuracy of the pure nonparametric method (VNP) and the nonparametric prior on the correction factor $\mathbf{C}$ (VNP-P) in \eqref{eq:correction_factor} that uses the spectral density matrix $\S_\text{p}$ of a best-fitting VAR(1) model. We consider three different block lengths 256, 512, and 1024, for the simulated time series from the VAR(2) model
\begin{align*}
    \Z_{t}=\begin{pmatrix}
        0.5 & 0\\
        0 & -0.3
    \end{pmatrix}\Z_{t-1} + \begin{pmatrix}
        0 & 0\\
        0 & -0.5
    \end{pmatrix} \Z_{t-2} + \bepsilon_{t}
\end{align*} 
for independent 
\begin{align*}
\displaystyle\bepsilon_t\sim N_2\left(\begin{pmatrix}
    0\\
    0
\end{pmatrix}, \begin{pmatrix}
    1 & 0.9\\
    0.9 & 1
\end{pmatrix} \right)
\end{align*}
and from the VMA(1) model
\begin{align*}
    \Z_{t}=\bepsilon_{t}+\begin{pmatrix}
        -0.75 & 0.5\\
        0.5 & 0.75
    \end{pmatrix}\bepsilon_{t-1}
\end{align*} 
for independent
\begin{align*}
\displaystyle\bepsilon_{t}\sim N_{2}\left(\begin{pmatrix}
    0\\
    0
\end{pmatrix}, \begin{pmatrix}
    1 & 0.5\\
    0.5 & 1
\end{pmatrix}\right).
\end{align*}


For each analysis, we ran a Markov chain of length \num{80000} and discarded the first \num{30000} iterations as burn-in. The remaining samples were thinned by a factor of 5, yielding \num{10000} posterior samples for subsequent inference.

\begin{figure}
    \includegraphics[width=1\columnwidth]{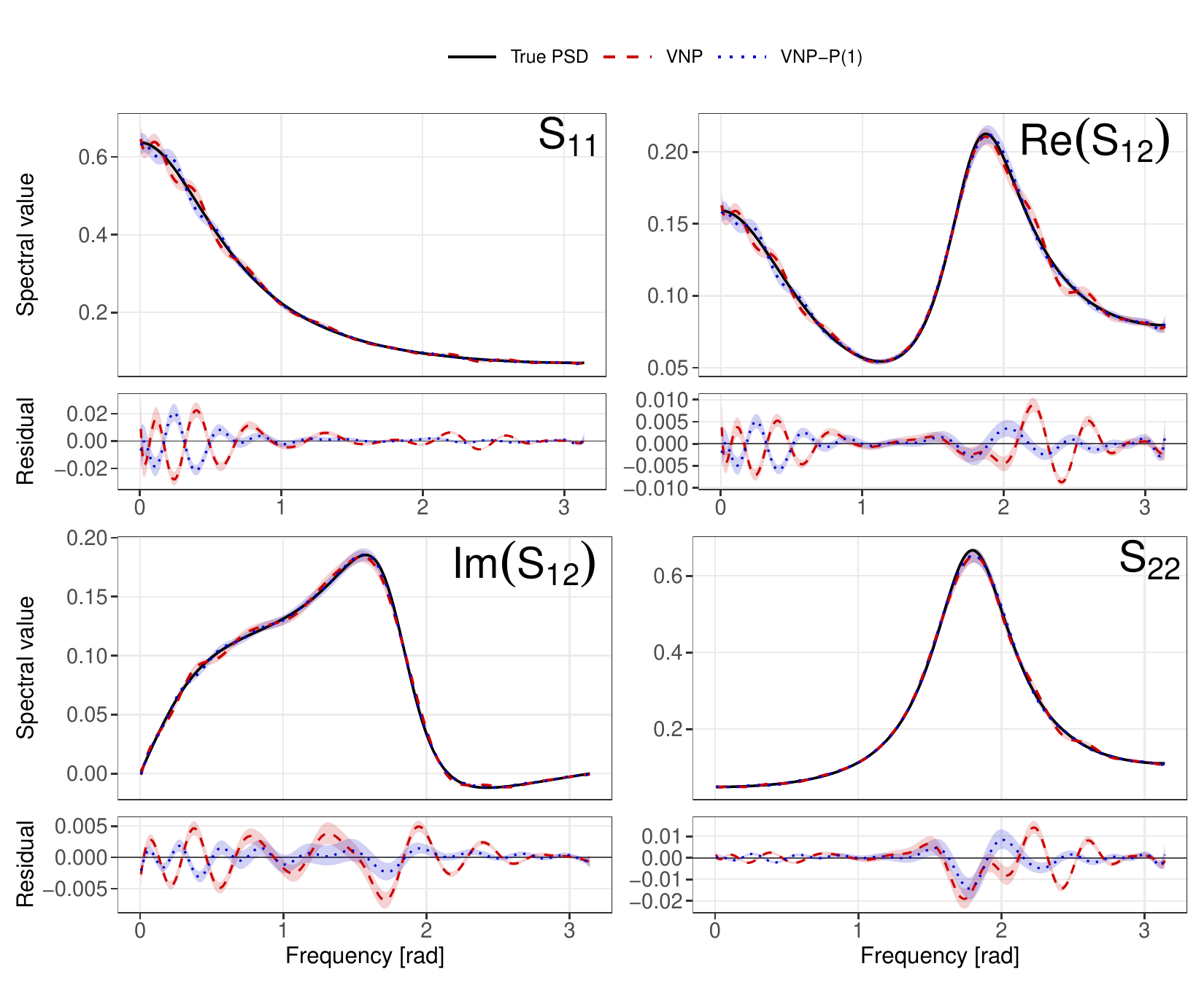}
    \caption{Median estimates for the VAR(2) model from VNP (red dashed line) and VNP-P (blue dashed line) for a randomly selected instance with a block length 1024. The true PSD is represented by a black solid line. The $90\%$ pointwise credible interval of VNP-P is shaded in blue. The residual plot for each PSD component is stacked below its corresponding estimate plot.
    }
    \label{fig:simulation_VAR(2)}
\end{figure}

Table~\ref{table:sim} compares posterior median PSD estimates in terms of $L_2$ error, frequentist coverage of the 90\% posterior credible interval, widths of the credible intervals for auto- and cross-spectra, and computation time. With increasing block length and fixed overall time-series length $n$, the estimation accuracy is expected to improve due to better frequency resolution, provided the accuracy of the averaged periodogram remains constant. Here, however, the accuracy of the averaged periodogram decreases because fewer segments contribute to the average.
For the VAR(2) case, the $L_2$ errors decrease only slightly  for both VNP and VNP-P, coverage and widths of credible intervals remain more or less constant. VNP-P achieves a smaller $L_2$ error than VNP by using a parametric VAR(1) model that can already model a significant part of the underlying correlation structure of the VAR(2) times series.  But this comes at the cost of increased computation time. Figure \ref{fig:simulation_VAR(2)} demonstrates the VNP and VNP-P estimates for one instance of  the VAR(2) time series with a block length of 1024.

For the VMA(1) case, the $L_2$ errors, coverage and widths of credible intervals remain stable with increasing block length. Here, using a VAR(1) parametric working model that cannot model the VMA(1) correlation structure  achieves no gain in accuracy compared to VNP. While the frequentist coverage of the credible intervals in the VMA(1) case -- even for the relatively small block lengths considered here -- is close to the nominal coverage, the empirical coverage for VAR(2) does not reach the nominal coverage yet for these finite block lengths.  It is well known that in infinite-dimensional models, posterior consistency does not guarantee that Bayesian credible intervals have the corresponding frequentist coverage \cite{Freedman1999}. A consistent posterior can still contract at the wrong rate and lead to poor coverage. Only for parametric models under certain regularity conditions, the Bernstein von Mises theorem guarantees that asymptotically credible intervals and frequentist confidence sets coincide \cite{Freedman1999}. 
However, as demonstrated by \cite{KnapikB.T.2011BIPW}, Bayesian credible sets typically have good frequentist coverage in case
of undersmoothing, i.e., when using a prior for the function to be estimated that is less regular than the truth. In our simulation studies with Bernstein-Dirichlet process priors for univariate spectral density estimation,  we have observed good frequentist coverage of Bayesian credible intervals \cite{KirchClaudia2019BwNc}.

\begingroup
\begin{table*}[!htbp]
\centering
\renewcommand{\arraystretch}{1.5}
\setlength{\tabcolsep}{3pt}
\begin{NiceTabular}{l l *{3}{|cc}}[colortbl-like, cell-space-limits=1pt]
\CodeBefore
  \rowcolors{5}{gray!10}{white}
  \columncolor{white}{1-2}
\Body
\Block{2-8}{} \\
  &   & \Block{1-2}{\bfseries$n/N_{b}=256$} & & \Block{1-2}{\bfseries $n/N_{b} = 512$} & & \Block{1-2}{\bfseries $n/N_{b} = 1024$} & \\
& & VNP-P & VNP & VNP-P & VNP & VNP-P & VNP \\
\hline
\Block{7-1}{\bfseries VAR(2)}
& $L_2$ error & 0.308 & 0.437 & 0.305 & 0.416 & 0.304 & 0.412 \\
& Pointwise coverage & 0.763 & 0.673 & 0.762 & 0.670 & 0.747 & 0.664 \\
& $\S_{11}$ CI Width & 0.071 & 0.063 & 0.073 & 0.063 & 0.072 & 0.064 \\
& $\Re \S_{12}$ CI Width & 0.069 & 0.067 & 0.068 & 0.066 & 0.068 & 0.066 \\
& $\Im \S_{12}$ CI Width & 0.061 & 0.058 & 0.062 & 0.058 & 0.062 & 0.058 \\
& $\S_{22}$ CI Width & 0.090 & 0.087 & 0.090 & 0.087 & 0.090 & 0.086 \\
& Time [min] & 21.00 & 14.52 & 41.17 & 30.87 & 84.95 & 48.67  \\
\hline 
\Block{7-1}{\bfseries VMA(1)}
& $L_2$ error & 0.227 & 0.240 & 0.276 & 0.258 & 0.332 & 0.294 \\
& Pointwise coverage & 0.920 & 0.897 & 0.919 & 0.882 & 0.895 & 0.877 \\
& $\S_{11}$ CI Width & 0.136 & 0.132 & 0.138 & 0.133 & 0.137 & 0.132 \\
& $\Re \S_{12}$ CI Width & 0.106 & 0.104 & 0.108 & 0.103 & 0.109 & 0.108 \\
& $\Im \S_{12}$ CI Width & 0.091 & 0.087 & 0.094 & 0.088 & 0.094 & 0.089 \\
& $\S_{22}$ CI Width & 0.205 & 0.195 & 0.207 & 0.197 & 0.208 & 0.205 \\
& Time [min] & 20.40  & 14.37 & 37.13  & 30.17 & 68.97  & 43.25 \\
\end{NiceTabular}
\caption{
Comparison of median $L_2$ errors, empirical pointwise coverage, median width of pointwise 90\% credible intervals, and median computation time for 500 simulations with the same full length, $n=\num{819200}$, using VNP-P and VNP  methods across different block sizes ($n/N_{b}=256$, $512$, and $1024$) for both VAR(2) and VMA(1) models.}
\label{table:sim}
\label{table:simstudy}
\end{table*}
\endgroup



\section{Applications}
We illustrate the use of the nonparametric VNP model on realistically simulated noise of the LISA detector and use ET-like noise to demonstrate the advantage of using an initial well-fitted parametric PSD when trying to detect correlated peaks in the ET spectrum. 

\subsection{Application to LISA}\label{subsec:LISA}
For the LISA detector, we use simulated instrumental-noise time series for second-generation time-delay interferometry (TDI) channels from the public noise-4a data release~\cite{noise4a}. The data set is noise-only, i.e., does not contain any gravitational wave signal, and uses the LDC Spritz instrumental-noise curves with constant, equal inter-spacecraft light-travel times of 8.3 seconds. The resulting time series includes the fixed-delay TDI response to the simulated instrumental noise, without orbital breathing or time-dependent arm-length modulation.
We apply the blocked Whittle likelihood to a one-year simulated LISA data set, sampled at 0.2 Hz, after segmenting the time series into 384 equal-length blocks of length \num{16384} data points, i.e., about 1-day blocks. 
A Hann window was applied to each segment to reduce spectral leakage. The MCMC sampler ran for \num{80000} iterations, and \num{70000} of them were discarded during the burn-in period and the remaining iterations were thinned by 2.

\begin{figure*}
    \centering
    \includegraphics[width=\textwidth]{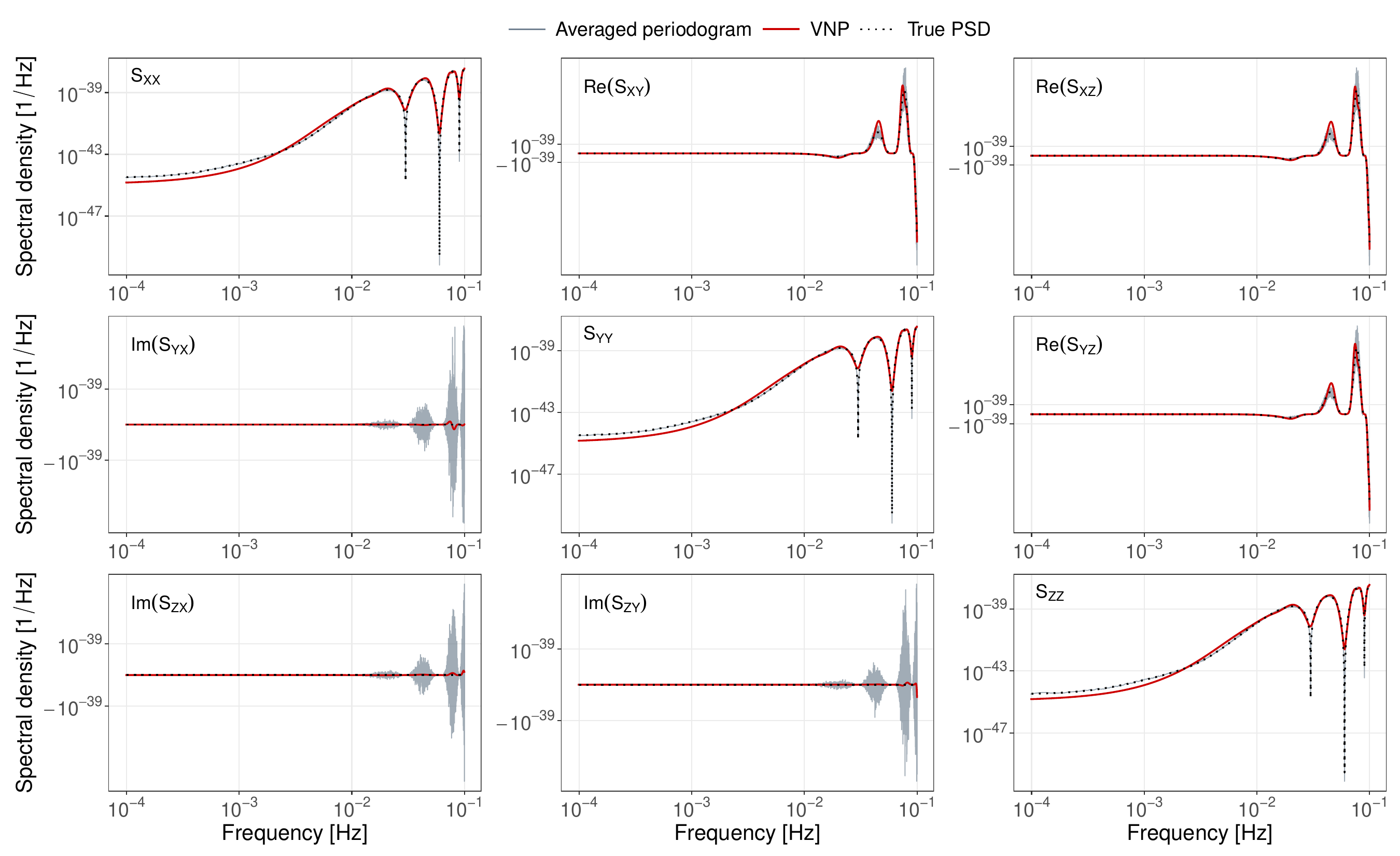}
    \caption{Median VNP estimates of the PSD of the simulated LISA noise are shown in red. The true PSD is represented by a black solid line, and the averaged periodogram is in gray.
    }
    \label{fig:lisa_psd_vnp}
\end{figure*}

\begin{figure}
    \centering
    \includegraphics[width=.9\columnwidth, height=0.5\textheight]{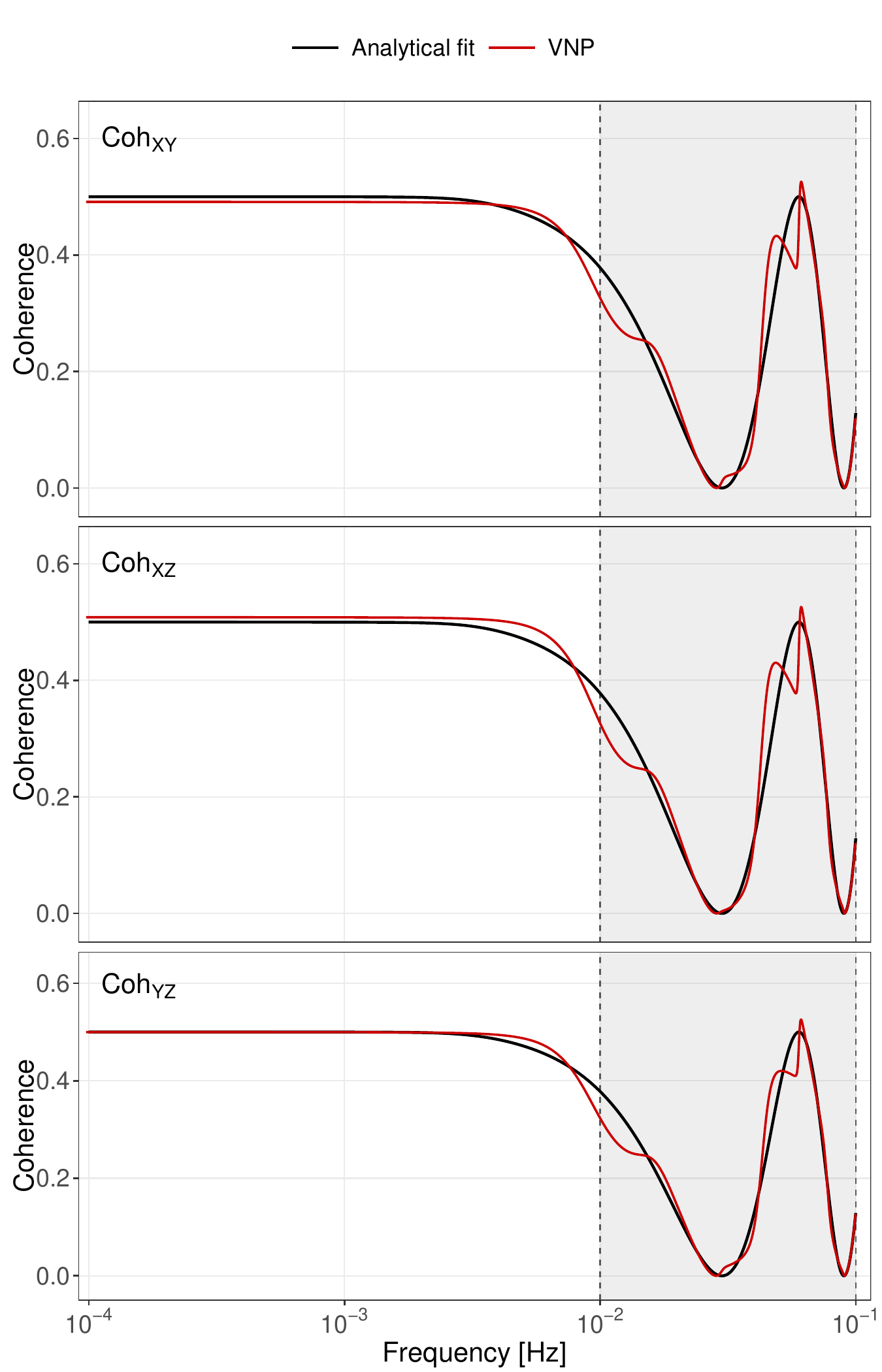}
    \caption{Median VNP estimates of the coherence of the simulated LISA noise is shown in red. The analytically squared coherence is presented in black.
    \avi{I think this plot can be improved by: 
    (5) the text
    discusses the transfer-function dips at $10^{-2}$--$10^{-1}$\,Hz -- it would
    help to shade or mark that region here (dashed vertical line maybe), since that is exactly where VNP
    departs from the truth and the caption gives the reader no hint that the
    deviation is expected.}}
    \label{fig:lisa_coh_vnp}
\end{figure}

Figure~\ref{fig:lisa_psd_vnp} compares the resulting nonparametric spectral estimate, obtained as the median of posterior MCMC samples, with the true PSD and averaged periodogram matrix. Figure~\ref{fig:lisa_coh_vnp} shows the corresponding coherence estimates. Overall, the spectral density estimate recovers the true PSD well. There is a slight underestimation at low frequencies, which might be caused by spectral leakage and could be suppressed by more aggressive tapering of individual blocks. The dips observed at frequencies $10^{-2}$ -- $10^{-1}$ Hz arise from the frequency-dependent transfer function, whose finite-arm-length response introduces response minima at specific frequencies \cite{TintoMassimo2021Ti}. In simulated TDI data, the strongly suppressed power near these frequencies can lead to increased relative fluctuations in PSD estimates and coherences, so apparent features in the vicinity of the dips should be interpreted with care. 

\subsection{Application to ET Noise}\label{subsec:ET}

Three independent Gaussian noise realizations were generated to represent the X, Y, and Z detector channels. Each realization has a duration of \SI{2000}{\second} and was filtered to reproduce the target power spectral density corresponding to the ET xylophone design sensitivity~\cite{Hild_2009,Hild:2010id}
\footnote{The sensitivity model adopted here corresponds to the former ``ET-D'' design sensitivity. We retain this version instead of the more recent sensitivity presented by ~\citet{Branchesi:2023mws} to facilitate direct comparison with the analysis of \citet{Janssens2023}. The differences between the two sensitivity curves are sufficiently small that they do not affect the conclusions of this work.}. The simulated data were sampled at \SI{2048}{Hz}, producing multichannel time series containing \SI{4096}{\kilo} samples in each detector channel. Throughout this paper, these realizations are collectively referred to as \textit{ET noise}.

To model correlated instrumental disturbances, we follow the approach of \citet{Janssens2023}. Additional colored Gaussian noise is injected into the X, Y, and Z channels, where each component is constructed from Gaussian-shaped features in the frequency domain. The corresponding power spectral density is

\begin{equation}
    \S_n^{GP}(f, \mu, A) = \left( \frac{A}{\sqrt{2\pi}} \exp\left\{ -\frac{1}{2}  (f-\mu)^2 \right\} \right)^2\, ,
\end{equation}
where $A$ represents the amplitude, and $\mu$ the frequency peak location.

Gaussian noise is injected at specific frequencies in each channel as follows
{\small 
\begin{align*}
    \textbf{X}&: \textstyle{\S_n^{GP}(\mu\eqq\SI{10}{Hz}, A\eqq4\ex{24}), \S_n^{GP}(\mu\eqq\SI{50}{Hz}, A\eqq2\ex{24})}\, ,\\
    \textbf{Y}&: \S_n^{GP}(\mu\eqq\SI{10}{Hz}, A\eqq4\ex{24}), \S_n^{GP}(\mu\eqq\SI{90}{Hz}, A\eqq1.5\ex{24})\, ,\\
    \textbf{Z}&: \S_n^{GP}(\mu\eqq\SI{50}{Hz}, A\eqq2\ex{24}),  \S_n^{GP}(\mu\eqq\SI{90}{Hz}, A\eqq1.5\ex{24}).
\end{align*}
}
To introduce correlated noise between channel pairs, we utilize identical Gaussian peaks at matching frequencies across the paired channels. 
%
This dataset is purposely simplified to facilitate the identification of the correlated noise terms \cite{Janssens2023} and serves as an initial demonstration of our nonparametric MCMC-based approach. In future work, this could be applied to more realistic and complex ET noise scenarios, including correlated magnetic noise \citet{Janssensmagneticnoise,Ball_lightning_strokes} and Newtonian noise \cite{Janssensnewtonianseismic,PhysRevD.109.102002}.
 
\begin{figure*}
    \centering
    \includegraphics[width=.85\textwidth,height=.26\textheight]{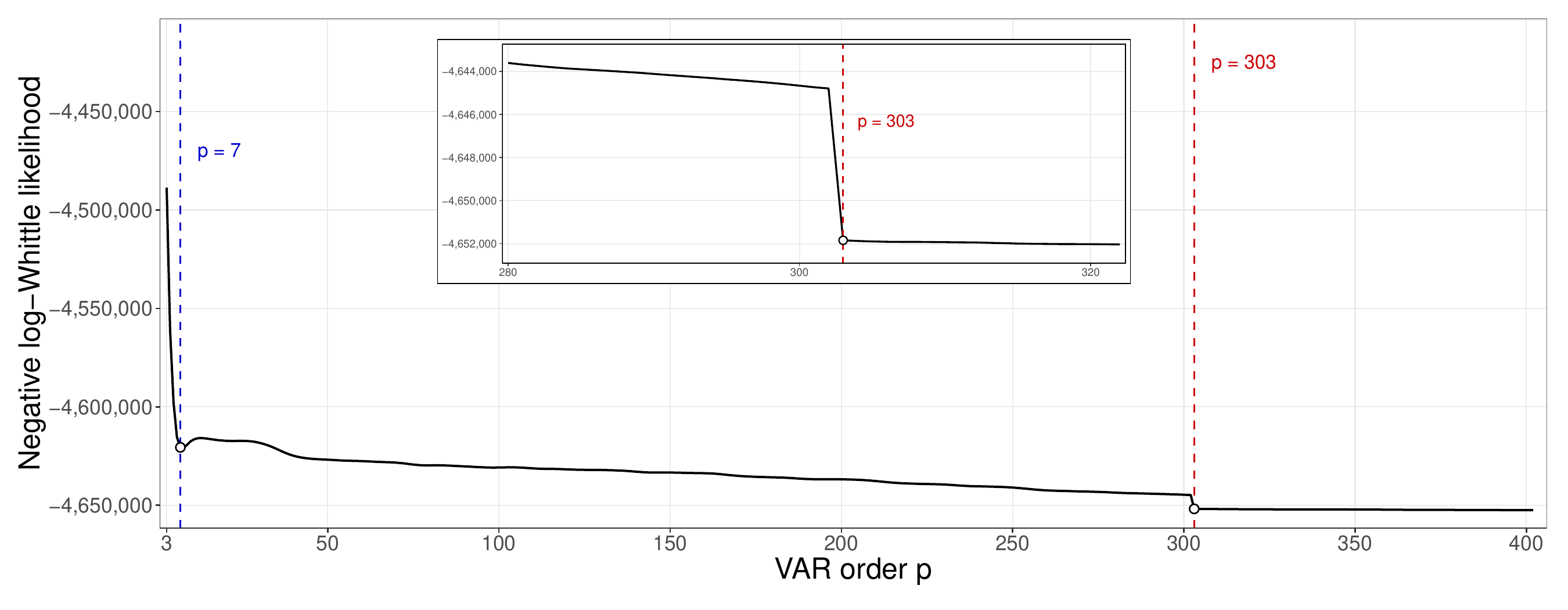}
    \caption{Negative log-Whittle-likelihood for prefitted VAR models with different autoregressive orders. The inset plot zooms in the orders around 303.
    }
    \label{fig:et_elbow_criterion}
\end{figure*}

The data is divided into 125 equal blocks, each consisting of \num{32768} data points and a maximum frequency of \SI{1024}{Hz} for the blocked Whittle likelihood estimation (Equation~\ref{eq:block_lnl}). No window function is applied to the data. The estimate focuses on the frequency range $5 - 128$ Hz.
Apart from the nonparametric VNP estimate, we also compute the VNP-P estimate. To this end, we propose to choose a parametric working model by fitting vector autoregressive models (VAR($p$)) of increasing order $p$ and choosing an adequate order using the so-called {\em elbow criterion}~\cite{KirchClaudia2019BwNc}, i.e., the order where a plot of autoregressive order $p$ versus the negative log-Whittle-likelihood displays an elbow or a significant dip. Figure \ref{fig:et_elbow_criterion} displays such an elbow plot 
for increasing autoregressive orders. 
The elbow criterion suggests order 7, but because the correlated peaks are small relative to the broadband noise, the log-likelihood shows a second large drop at 303. A VAR(7) model does not capture the peaks.
We therefore select order 303 for the parametric VAR fit. For both VNP and VNP-P approaches, we ran \num{14000} iterations and discarded \num{12000} of them during the burn-in period, without thinning for the remaining iterations.

\begin{figure*}
    \centering
    \includegraphics[width=.9\textwidth,height=.6\textheight]{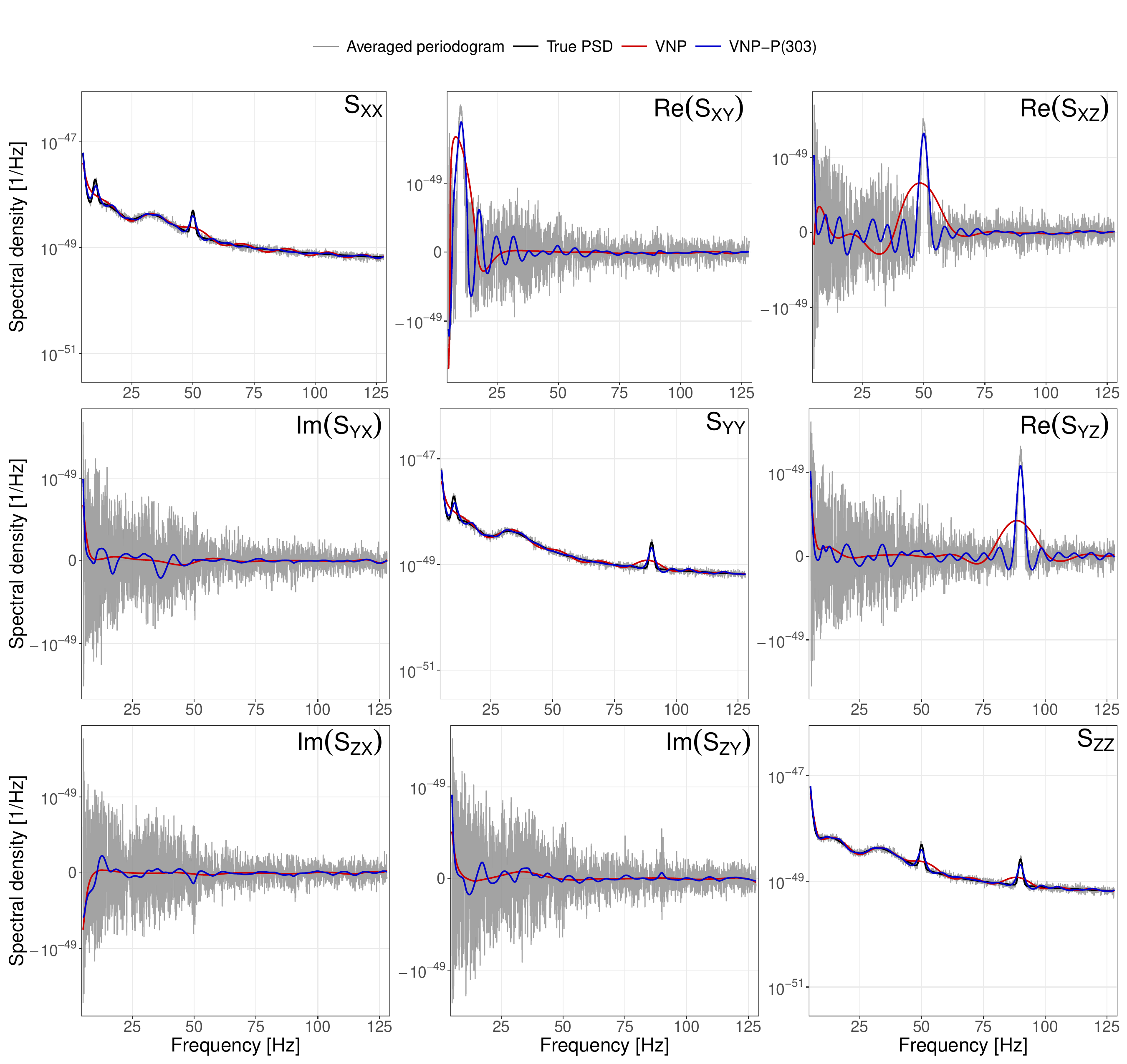}
    \caption{Median estimates of the PSD of the ET data from VNP (red) and VNP-P(303) (blue). The periodogram is represented in gray, and the true PSD is in black. 
    \avi{I think this plot can be improved by: (1) exporting it as a PDF -- it is
    the only raster figure in the paper (a 5400$\times$3000 PNG), so it will look
    softer than the others in print and it bloats the source; (2) the caption
    says VNP-P is ``blue'' but it is drawn in \emph{cyan}, which is also a
    different colour from the blue used for VNP-P in
    Fig.~\ref{fig:simulation_VAR(2)} -- please fix one way or the other (see my
    general comment) (3) the light-grey periodogram fills essentially the whole
    panel and swamps the curves we actually care about -- either plot the
    \emph{block-averaged} periodogram (which is what the likelihood uses) instead
    of the raw one, or reduce its alpha a lot; (4) the correlated peaks at 10,
    50 and 90\,Hz are the entire point of this figure and they are barely
    visible -- I would add zoomed insets (or a second row of panels) around each
    peak, otherwise the reader has to take the ``VNP oversmoothes, VNP-P captures
    them'' claim on trust (5) same $y$-axis notation issue as
    Fig.~\ref{fig:lisa_psd_vnp} (log ticks on the diagonal,
    \texttt{1e-49}-style on the off-diagonals) and same missing units; (6) the
    legend sits inside the bottom-right panel over the data -- move it out and
    (7) ``VNP-P(303)'' in the legend vs ``VNP-P'' in the caption -- please make
    the legend entries identical across all figures.}}
    \label{fig:et_psd_vnp_VNP-P}
\end{figure*}

\begin{figure}
    \centering
    \includegraphics[width=1\columnwidth]{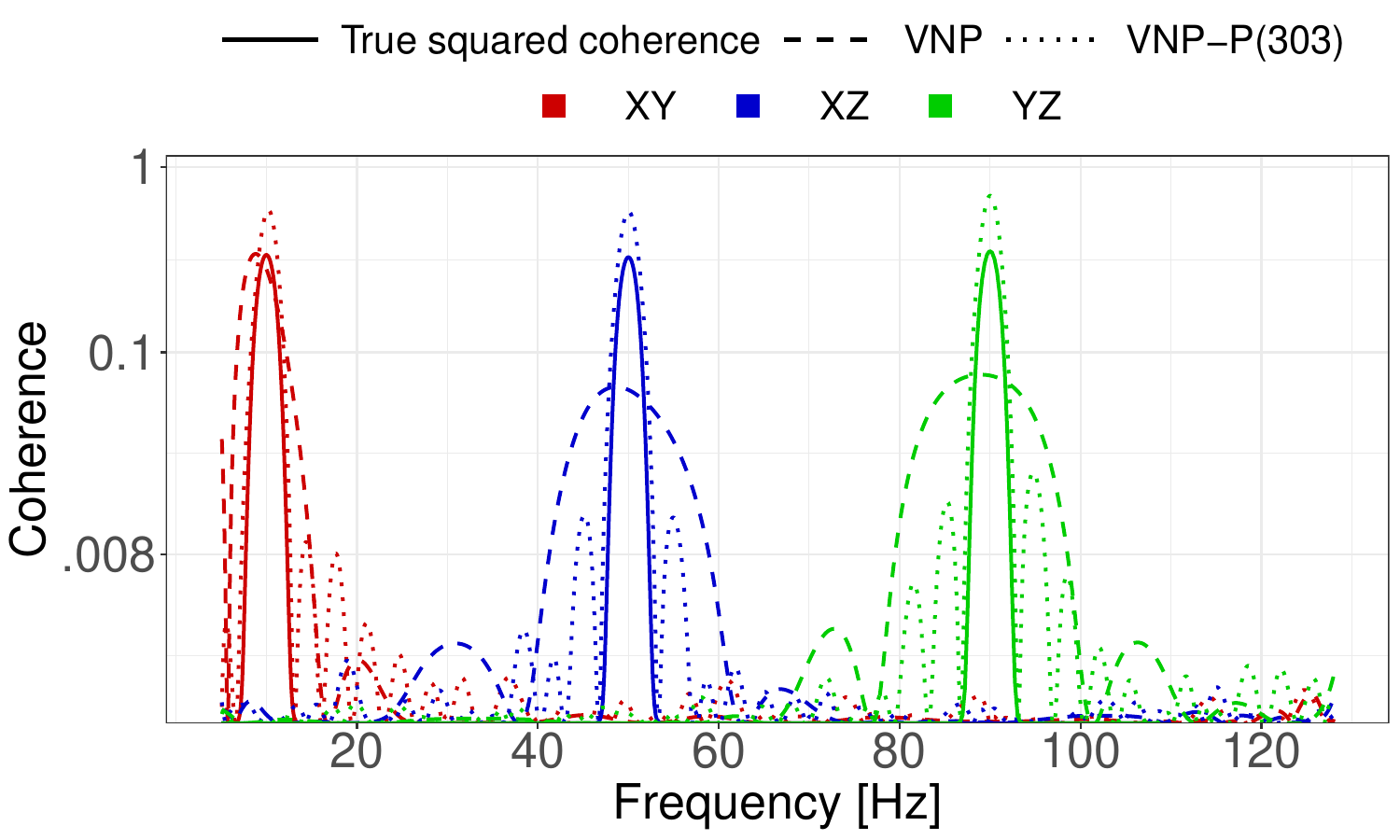}
    \caption{Median estimates of the squared coherence of the ET data from VNP (dashed) and VNP-P(303) (dotted) with an order of 303. True coherence is represented by a solid line. 
}
    \label{fig:et_coherence_vnp_VNP-P}
\end{figure}

Figure \ref{fig:et_psd_vnp_VNP-P} shows the periodogram matrix overlaid by the median of posterior samples of VNP and VNP-P. Here we see the benefits of using a parametric working model to improve the performance of the nonparametric method. While VNP oversmoothes the correlated peaks, VNP-P clearly captures them. This finding is further supported by the squared coherence estimates shown in Figure~\ref{fig:et_coherence_vnp_VNP-P}. Although VNP accurately reconstructs the coherence peak at 10 Hz, its estimates at 50 Hz and 90 Hz are oversmoothed. By contrast, VNP-P successfully captures all three peaks. VNP-P fits the simulated ET data better than VNP, achieving an $L_2$ error of 0.068 versus 0.078 for VNP. The running time for VNP is 24.478 mins, compared with 33.421 mins for VNP-P. 


\section{Discussion}\label{sec:discussion}
This paper presents a flexible nonparametric Bayesian approach (VNP) for estimating the spectral density matrix of correlated time series and demonstrates its performance through  a simulation study with two data-generating processes, as well as simulated correlated noise from LISA and ET. The method addresses a challenging problem in gravitational-wave data analysis for next-generation detectors such as ET and LISA, where correlations between multiple data channels must be properly accounted for to avoid biased gravitational-wave parameter estimation.

The flexible prior model used for the spectral density is based on a Bernstein polynomial basis expansion where matrix-valued weights are modelled nonparametrically using a matrix-gamma process.
To accommodate the extremely long time series encountered in applications such as LISA, which may contain many thousands of observations, we divide the data into segments and average the corresponding periodogram matrices. Under the assumption that the segments are independent, we show that the resulting blocked likelihood reduces to a product of independent complex Wishart distributions. We further introduce a semiparametric extension that incorporates a rough parametric spectral estimate while remaining robust to model misspecification through a nonparametric correction term (VNP-P). Rather than modeling the spectral density matrix itself nonparametrically, VNP-P models a correction function that captures deviations from the parametric working model.

The VAR(2) simulations suggest that the results are not highly sensitive to the range of block lengths considered here. Increasing block lengths from 256 to 1024 produces only small changes in $L_2$ error, pointwise coverage, and credible-interval width. The parametric-correction model improves accuracy, VNP-P has lower median $L_2$ error than VNP for all three block lengths, while pointwise coverage remains similar. This supports the intended role of the parametric working model as a rough spectral template that captures part of the broad VAR structure, with the nonparametric component modeling the residual discrepancy.
In contrast, simulations based on a VMA(1) process show that VNP-P does not improve accuracy when the parametric working model is misspecified as a VAR(1) process. This suggests that the benefits of the parametric-correction approach depend on the ability of the working model to capture at least some of the dominant spectral features of the underlying process.

We demonstrated the nonparametric VNP approach with an application to simulated LISA TDI data and successfully recovered the true underlying PSD and the coherences, even in the high frequency range. The application to correlated ET channels highlighted the benefit of employing a parametric working model that already captures the main features of the periodogram matrix and models the relative deviations from this using the nonparametric matrix-gamma approach. Guided by the parametric model, VNP-P was able to capture the correlated peaks, whereas the pure nonparametric 
VNP method oversmoothed the peaks. With an $L_2$ error of $0.068$ compared with $0.078$ for VNP, VNP-P provides a better fit to the simulated ET data.


Although the VNP-P approach generally outperforms VNP when a suitable parametric model is available, this improvement comes at the expense of increased computational cost. \textcolor{black}{The order $p$ of the VAR working model controls the flexibility of the parametric model and therefore needs to be carefully chosen to avoid overfitting.  Here, we used the {\em elbow criterion} which has been shown to work well in practice and is fast to compute based on maximizing the Whittle likelihood \cite{KirchClaudia2019BwNc}. A more computationally involved Bayesian criterion by \cite{Blinks} estimates the VAR order. Their approach is based on a reparametrization of the VAR model in terms of partial autocorrelation matrices and putting a multiplicative gamma  prior on $p$ that encourages shrinkage of the partial autocorrelations with increasing lag. Posterior inference is based on Hamiltonian Monte Carlo and would add another layer of computational complexity. However,
as demonstrated in \cite{KirchClaudia2019BwNc}, the parametric model does not need to provide a very close fit but should leave room for a nonparametric correction to provide robustness against a mis-specified parametric model. It would still be interesting to explore the relative performance of the approach by \cite{Blinks} compared to the elbow criterion in terms of both accuracy and robustness. We leave this for future research.}

Another limitation is that the LISA and ET applications considered in this work are restricted to noise-only data. An important direction for future research is to incorporate our methodology into a Gibbs sampling framework for joint signal-plus-noise (on-source) estimation, which is particularly relevant for LISA analyses. Another promising avenue is to extend the proposed approach to more realistic and complex ET noise scenarios, including correlated magnetic noise \cite{Janssensmagneticnoise,Ball_lightning_strokes} and Newtonian noise \cite{Janssensnewtonianseismic,PhysRevD.109.102002}, and realistic LISA noise containing gaps and glitches.



\section*{Data and Software Availability}
The R-package developed for this research is available on GitHub at \url{https://github.com/easycure1/vnpc.avg}. The copy of the R-package and the data for this work have been archived on Zenodo \cite{mcmc_zenodo}. The datasets generated and analyzed during the current study are available from the corresponding author upon reasonable request.

\begin{acknowledgments}
We thank Jean-Baptiste Bayle for providing the LISA dataset 
and Kamiel Janssens for generating the correlated ET noise data.
 AV,  JEL, NC, PMR and RM gratefully acknowledge support  by the Marsden Fund Council grant MFP-UOA2531 from New Zealand Government funding, managed by the Royal Society Te Apārangi. 
 This work was supported by the CNRS International Research Project (IRP) OG-Science FR-NZ. NC has received financial support from the French Agence Nationale de la Recherche. We thank the New Zealand eScience Infrastructure
(NeSI) \url{https://www.nesi.org.nz} for the use of their high performance computing facilities and the Centre for eResearch at the University of Auckland for their technical
support.
\end{acknowledgments}
\bibliography{reference}


\end{document}